# VOID PREDICTION DURING LIQUID COMPOSITE MOLDING PROCESSES: WETTING AND CAPILLARY PHENOMENA

A. Ben Abdelwahed[1], Y. Wielhorski[1], L. Bizet[1], J. Bréard[1*]

[1]Laboratoire Ondes et Milieux Complexes (LOMC), UMR 6294 CNRS, Université du Havre, 53 rue de Prony 76058 Le Havre, France
*joel.breard@univ-lehavre.fr

**Keywords:** LCM, Bubble, Pore Doublet Model, Millifluidic.

**Abstract**
*The aim of this work is to contribute in improving fibrous preforms impregnation for Liquid Composite Molding (LCM) processes. The void prediction in LCM sparks off interest within the Composite Material elaboration because it represents a significant issue to keep the expected mechanical properties of the final product. The liquid properties, the preform geometry and the flow conditions impact the void or bubble entrapped inside and outside the yarns. Nevertheless, due to the complex geometry of the reinforcement, experimental characterization of bubble formation remains delicate. Thus, our study deals with two simple model networks representing connected pores so called "Pore Doublet Model". A first is considering two capillaries converging on a node (T-junction) and a second is representing two capillaries interconnected with a supplying principle. In this paper, we emphasize on microfluidic and millifluidic approaches where wetting and capillary forces are significant during bubble formation mechanism.*

## 1 Introduction

*1.1 Bubble formation during Liquid Composite Molding processes*
The LCM processes refer to composite manufacturing techniques where a resin is injected through a fibrous preform with an imposed pressure or flow. During these processes, bubbles can be created and then entrapped. The origins of this phenomenon stem mainly from the complex geometry of the reinforcement and the adhesion properties at the liquid/fibre interface. The fibrous preform is often represented with mainly two pore scales: a macroscale between the bundles and a microscale within the bundles. Thus, the flows are governed differently according to the scale, either by the capillary pressure inside the tow or by the viscous force between the tows. Hence, the bubble formation can be explained by the competition between the capillary and the viscous effects which are compared by the capillary number Ca. Indeed, these both kinds of flow induce locally a difference between the front positions leading consequently to the bubble formation that can be summarized as follows: the intra-tow void, so-called microvoid, which is located inside the bundle and the inter-tow void, also called macrovoid, between two consecutive bundles (Figure 1). More precisely, the microvoids occur at high capillary number Ca when the viscous flow overcomes the capillary forces whereas the macrovoids are obtained at lower Ca where the capillary pressure overcomes the viscous pressure.

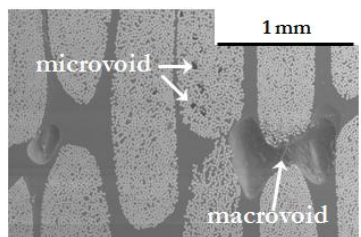

**Figure 1.** Microvoid and macrovoid entrapment inside a fibrous perform.

To quantify the void and attempt to minimize its rate inside composite materials, many experimental studies have been carried out [1-4]. However, it is often difficult to visualize bubble formation and transport during the LCM process. Therefore, to determine the final saturation, numerical works [5], experimental and theoretical studies in porous Pore-Doublet Model [4, 6] are ways to get round this difficulty. The latter is a widely used modelling to understand the dynamic of immiscible fluids in porous media [7, 8].

*1.2 Modelling bubble formation in connected pores*
First, the purpose of this paper is to underline the influence of the wetting situation when generating a confined bubble train using a technique based on interaction between two immiscible fluids (liquid-gas). Experiments were made using a cylindrical T-junction device, widely employed for the cross-flowing configuration, which can model two convergent pores from the fibrous preform. Secondly, we emphasise the supplying principle arising from the bulk provided by the macrochannel, which plays the role of a tank, to the microchannel [4, 6, 16]. A theoretical model for a wetting situation will be presented by taking into account supplying principle within interconnected capillaries.

*1.3 Bubble formation by cross-flowing streams: Microfluidic approach*
In the last decade, the interest in the interactions between to immiscible fluids (liquid/liquid or liquid/gas) by converging a dispersed phase (break-up stream) and a continuous phase (shear stream) has risen significantly. Indeed, several microsystems were developed and many studies have been carried out and extended to multiphase flow, in order to improve the industrial microfluidic devices employing calibrated bubbles and droplets. In this respect, some of the main investigations have been reviewed lately [9] focusing on the physical mechanisms governing the bubble creation in various microfluidic devices. Furthermore, three fundamental techniques have been employed to form droplets and bubbles: *i)* the co-flowing streams [10], *ii)* the cross-flowing streams [11] and *iii)* the microfluidic flow focusing devices [12]. As for flows with low Reynolds numbers, the use of microchannels in the above-named systems allows to consider the viscous and interfacial effects when describing the dynamic formation of calibrated bubbles and droplets. The bubble generating system we used in our study is a T-junction device which is widely employed for the cross-flowing streams mechanism. It consists in converging a continuous liquid stream and a dispersed gas phase perpendicularly (Figure 2).

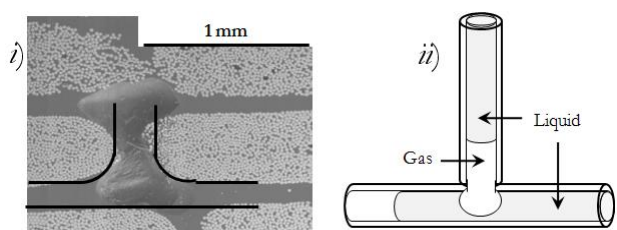

**Figure 2.** *i)* Composite material with an entrapped bubble and *ii)* its equivalent model sketch.

Two main regimes monitoring bubble formation can be distinguished according to the range of the capillary number values: the squeezing and the dripping regimes. The first is a confined breakup mechanism occurring at typically small capillary numbers (Ca < $10^{-2}$), meaning that the interfacial force overcomes the viscous shear stress [11]. The obtained bubbles, so called slug bubbles, are quite longer than the channel width. Two phases characterise the bubble growth [11]: the filling phase during which the dispersed phase penetrates into the continuous stream until the bubble becomes large enough to obstruct the main flow, and a squeezing phase during which the incoming liquid pushes the bubble downstream with a velocity close to the mean velocity of the liquid until the break-up. During the filling period, it was established that the bubble size doesn't depend on the flow conditions but is influenced by the width of both channels [13]. It was also found that the bubble length scales linearly with the ratio of the flow rates of both the dispersed and the continuous phases during the squeezing period [11]. The second regime is an unconfined breakup mechanism occurring when Ca ≥ 2 $10^{-2}$ [14]. It takes place when the dispersed phase doesn't clog the continuous phase and the bubble breakup is controlled by the local shear stress. In a recent work [15], we have found that the bubble size and velocity depends on the wetting behaviour between the injected liquid and the solid surface used. Indeed, for a non-wetting configuration, there was a discrepancy between the measured bubble length and the one expected by the models. Up to our knowledge, little attention has been paid to bubble formation involving partially wetting situations. We will focus on the bubble size in the squeezing regime only (for small Ca) in which the interfacial forces should be predominant.

## 2 Materials and testing methods

*2.1 Wetting characterization*

We used silicone polydimethysiloxane oils $(CH_3)_3Si - O - [(CH_3)_2SiO]_n - Si(CH_3)_3$ (PDMS Rhodorsil 47V100 from Rhodia), n-Hexadecane 99% $CH_3(CH_2)_{14}CH_3$, Ethylene glycol ≥ 99.5% $C_2H_6O_2$, distilled water $H_2O$ and Glycerol 98% $C_3H_8O_3$. The liquid surface tensions $\gamma_L$ and the apparent static contact angles $\theta_s$ on the glass tube were respectively determined by the Wilhelmy and the Jurin methods with a Krüss K100SF tensiometer. For each liquid, both $\gamma_L$ and $\theta_s$ were measured five times and the mean values was selected. The liquid properties (liquid surface tension $\gamma_L$, density $\rho$ and viscosity $\eta$) and the static contact angle measurements are given in Table 1.

|  | $\gamma_L \pm 0.4$ [mN.m$^{-1}$] | Literature $\gamma_L$ [mN.m$^{-1}$] | $\rho$ [g.cm$^{-3}$] | $\eta$ [mPa.s] | $\theta_s \pm 2$ [°] |
|---|---|---|---|---|---|
| 47V100 | 20.8 | 20.9 [a] | 0.965 [a] | 100.0 [a] | 13 |
| Hexadecane | 27.5 | 27.6 [b] | 0.773 | 3.3 | 34 |
| Ethylene glycol | 48.7 | 48.3 [c] | 1.109 | 21.8 | 77 |
| Water | 72.6 | 72.8 [b] | 0.998 | 1.0 | 89 |
| Glycerol 98% | 62.9 | 63.4 [b],[c] | 1.263 | 1100.0 | 90 |

**Table 1.** Liquid properties and wetting characterization at 20°C: [a] Rhodia Rhodorsil, [b] Ström (Krüss database) and [c] Fowkes (Krüss database)

*2.2 Bubble formation in T-junction device*

Experiments were performed in a cylindrical glass capillary tube with inner radius $R_c$ = 1.0 mm. Liquid and gas were injected with controlled volumetric flow rates thanks to syringe pumps: $Q_1$ represents the continuous liquid flow rate in the main channel and $Q_2$ is the flow rate of the gas discontinuous phase (Figure 3).

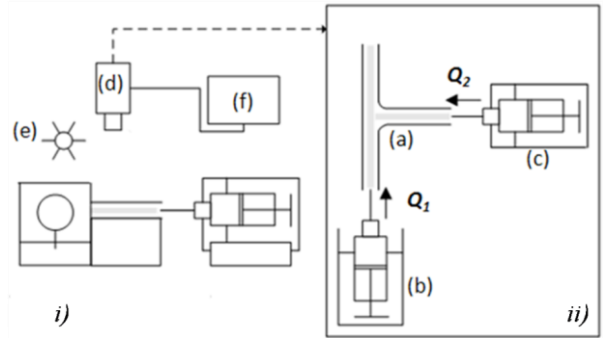

**Figure 3.** Simplified outline of the experimental device: *i)* side view; *ii)* top view.

Note that the range of the ratio $Q_2/Q_1$ is between $10^{-2}$ and 3. The injected gas penetrates into the main channel until the opposite wall and was sheared by the liquid stream. The bubble was pinched, broken up and carried away by the cross flow. After the change in the flow rate values, we waited an adequate time before taking measurements to let the system relax to a steady state flow and a stable break-up mechanism. The experimental study was performed by controlling the main dimensionless numbers, which are given in Table 2.

| Dimensionless numbers | | Range |
|---|---|---|
| Reynolds number | $Re_1 = 2\rho U_1 R_c / \eta$ | $1.4\ 10^{-4} \leq Re_1 \leq 5.0\ 10^{-1}$ |
| Capillary number | $Ca_1 = \eta U_1 / \gamma_L$ | $2.5\ 10^{-7} \leq Ca_1 \leq 1.0\ 10^{-2}$ |
| Bound number | $Bo = \rho g R_c^2 / \gamma_L$ | $1.3\ 10^{-1} \leq Bo \leq 4.5\ 10^{-1}$ |
| Froude number | $Fr_1 = U_1^2 / 2gR_c$ | $1.6\ 10^{-8} \leq Fr_1 \leq 1.4\ 10^{-3}$ |

**Table 2.** Dimensionless number ranges of the continuous phase.

The bubble formations were filmed and images were extracted and then analyzed by detecting the liquid front position and the bubble tip with an accuracy of about 40 µm. Before the experiment, the glass T-junction device was immersed for one day in a drenching of solution made with 98% of distilled water and 2% of Decon 90. Then, the capillary was flushed with distilled water and air and baked in the oven at 105°C for about two hours.

## 3 Results and discussion
*3.1 Bubble length in the squeezing regime*
Figure 4 shows the normalized bubble length as a function of the flow rate ratio $Q_2/Q_1$.

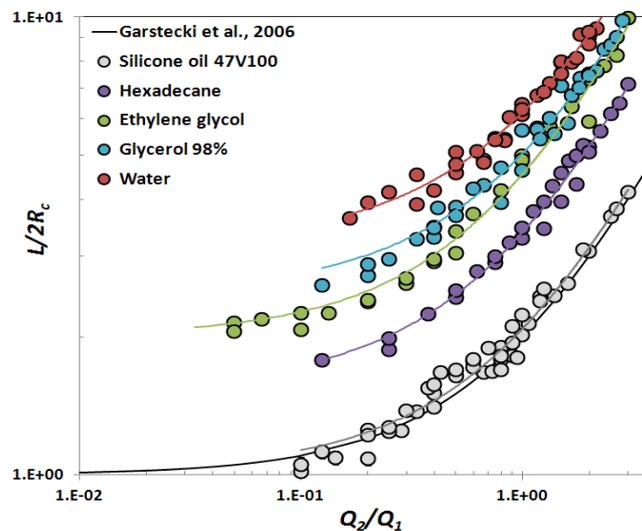

**Figure 4.** Normalized bubble length as a function of flow rate ratio.

The experimental trends give a linearly growth of the bubble length with $Q_2/Q_1$ that can be expressed as:

$$\frac{L}{2R_c} = A\frac{Q_2}{Q_1} + B \tag{1}$$

Here A and B are the fitted coefficients and are reported in Table 3.

| | A | B | $Q_2/Q_1 = 2$ |
|---|---|---|---|
| Silicone oil 47V100 | 1.06 ± 0.05 | 1.03 ± 0.10 | 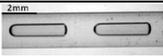 |
| Hexadecane | 1.83 ± 0.20 | 1.56 ± 0.20 | 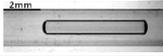 |
| Ethylene glycol | 2.52 ± 0.20 | 2.02 ± 0.20 | 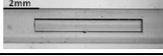 |
| Glycerol 98% | 2.52 ± 0.20 | 2.52 ± 0.20 | 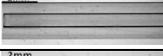 |
| Water | 2.97 ± 0.20 | 3.22 ± 0.20 | 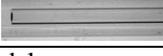 |

**Table 3.** Fitting coefficients for bubble length model.

The different values of A and B can be explained by depicting the pressures acting on the bubble during its formation: namely, the driving pressure exerted by the shear stream force, the surface tension force and the viscous dissipations arising from the viscous resistance in the core of the liquid, the dissipation within the precursor film and the friction in the contact line.
For the silicone oil, the results shown in Figure 4 and Table 3 indicate that the experimental coefficients A and B are close to those predicted by Garstecki et al. model's (A ≈ B ≈ 1) [11]. In this case, the measured apparent static contact angle $\theta_s$ is about 13°± 2°. The low value of the contact angle means that bubble formation occur in a quasi complete wetting configuration which is close to the above named studies [11, 13].
For the other liquids, the apparent contact angle $\theta_s$ is between 34° and 90° meaning that they are partially wetting liquids. The bubble length evolution remains linear but with experimental coefficients A and B higher than those acquired for the completely wetting case. For instance, the obtained coefficients for the Hexadecane are about 1.6 times higher than those for the silicone oil. For relatively higher apparent contact angles (77° ≤ $\theta_s$ ≤ 90°), the bubble lengths are about twice to three-fold larger than those for the wetting liquids. In the partial wetting case, we observe that the system bubble-liquid slug is less dissipative than in the wetting case. Indeed, the surface tension forces are weaker due to the increase of $\theta_s$ and the dissipation in the precursor film is not a prevalent feature. Thus, one should to take into account the dissipation due to the viscous flow and due to the frictional processes near the contact line. The latter is unstable due to the molecular interactions at the solid-liquid interface. This qualitative description can explain why the bubbles are larger than in the wetting case. Thereby, when the spreading energy is low (high contact angle), the contact line is more mobile and we can expect to have larger bubble lengths. Moreover, the uncertainties about the bubble size measured in the partial wetting situation are higher than for wetting liquids because the position of the contact line is not known in advance, this is free-boundary complex problem.

*3.2 Interconnected capillaries: theoretical approach*
The pore doublet model studied in this part is composed of two circular capillaries with different radii: $R_M$, for the larger capillary, called "macrochannel" and $R_m$ for the smaller one, named "microchannel". The PDM capillaries are divided into two parts over the length: a first part where the capillaries are continuously interconnected on over a distance $l$, so-called "continuous interconnectivity", subsequently followed by a second part of length $L$, wherein the capillaries are interconnected at both ends forming nodes, so-named "node interconnectivity" (Figure 5). Note that the void entrapped in the macrochannel and in the microchannel will be respectively called macrovoid and microvoid. The filling of the PDM can be divided into two phases according to the position x(t) of the both menisci into the two parts. The first phase, which is defined for the time range [0,$t_{fl}$], occurs when one of the both menisci reaches first the end of the first part (x($t_{fl}$) = $l$). The second phase, defined for $t_{fl} \leq t \leq t_{fL}$, is when one of the both streams completely fills the second part (x($t_{fL}$) = $l+L$). In order to simplify the following analytical development, the void is assumed to be incompressible and the liquid as Newtonian. Furthermore, we will consider that the advanced contact angle at the intersection between the liquid-gas interface and the solid surface is supposed to be approximately equal to the apparent equilibrium contact angle. To lighten the notations, we will use the subscripts $m$ and $M$, respectively for the microchannel and the macrochannel and the continuous interconnectivity and the node interconnectivity parts are noted with respectively the exponents $ci$ and $ni$. The governing flow equations describing the forces filling of the PDM is given below.

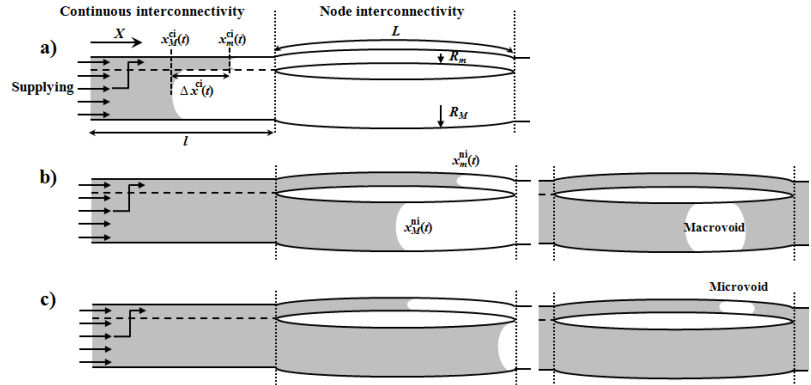

**Figure 5.** Sketch of void formation inside a PDM formed by a continuous interconnectivity and a node interconnectivity parts: a) Supplying principle, b) macrovoid and c) microvoid creation.

We consider the supplying principle [16] that consists in supplying mass to the microchannel from the macrochannel which plays the role of a tank. Consequently, the meniscus inside the microchannel is always ahead. The difference between the both menisci $\Delta x^{ci}(t)$, balancing the viscous pressure drop with the capillary pressure between the both menisci, reads:

$$\Delta x(t) = \left( \frac{\gamma_L \cos\theta_s (R_M - R_m)^2 R_m^2}{2\eta R_M^3} \right)^{1/2} t^{1/2} \qquad (2)$$

For the first part of the macrochannel and for the second part of the both capillaries, the motion equation is deduced from balancing the pressures, with the addition of an injection pressure $P_i$ to the Lucas-Washburn equation [17, 18]:

$$\frac{8\eta}{R_{m,M}^2} x \dot{x} = P_i + \frac{2\gamma_L \cos\theta_s}{R_{m,M}} \qquad (3)$$

Besides, we define a pressure $P^*$, for which the both menisci arrive at the same time $t_{fl}$ at the node $x = l$, expressed as $P^* = (2\gamma_L\cos\theta_s/R_M) [\alpha(2+\alpha-\alpha^2)/(1+\alpha)]$, where $\alpha \in [0;1]$ representing the ratio of capillary radii $R_m/R_M$. In the following, we set $\beta$ the part length ratio $L/l$.

For $Pi < P^*$, we assume that the meniscus in the small capillary reaches first the first node. Two possibilities can occur: *i)* the meniscus inside the macrochannel reaches first the second node, creating therefore the microvoid (Eq. 4); or *ii)* the stream inside the microchannel remains in advance and reaches first the second node, hence the macrovoid is created (Eq. 5).

$$\Delta x_m^{ni}(t_{fL}) = l \left[ \beta - \alpha \left( \frac{P_i R_m + 2\gamma_L \cos\theta_s}{P_i R_m + 2\alpha\gamma_L \cos\theta_s} \right)^{1/2} \left( 1 + \beta - \frac{x_M^{ci}(t_{fl})}{l} \right) \right] \quad (4)$$

$$\Delta x_M^{ni}(t_{fL}) = l \left[ 1 + \beta - \frac{\beta}{\alpha} \left( \frac{P_i R_m + 2\alpha\gamma_L \cos\theta_s}{P_i R_m + 2\gamma_L \cos\theta_s} \right)^{1/2} - \frac{x_M^{ci}(t_{fl})}{l} \right] \quad (5)$$

For $Pi > P^*$, we suppose that the flow inside the macrochannel reaches first the first node. We have only one possibility can be obtained and the entrapped microvoid reads:

$$\Delta x_m^{ni}(t_{fL}) = l \left[ 1 + \beta - \alpha\beta \left( \frac{P_i R_m + 2\gamma_L \cos\theta_s}{P_i R_m + 2\alpha\gamma_L \cos\theta_s} \right)^{1/2} - \frac{x_m^{ci}(t_{fl})}{l} \right] \quad (6)$$

Curves plotted in Figure 6.*a)* are given for α = 0.1 and by varying the parameter β. For each value of β and for $Pi < P^*$, the macrovoid rate decreases with the increase of the pressure Pi because the stream inside the macrochannel is flowing increasingly quickly. Besides, for a given Pi, the macrovoid rate grows with the decrease of β. Curves presented in Figure 6.*b)* are obtained for β = 2 10^{-3} and different values of α.

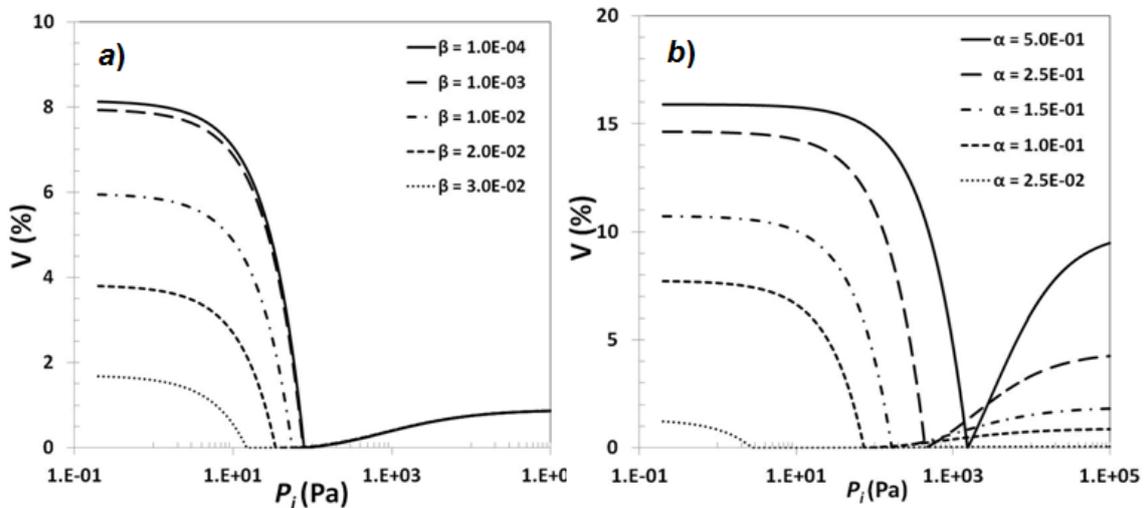

**Figure 6.** Void rate evolution as a function of the imposed pressure $P_i$ for $l$ = 500mm and $R_m$ = 10μm: *a)* Different values of *β* with *α* = 0.1 and *b)* different values of *α* with *β* = 0.002.

## 4 Conclusion

We have proposed issues to quantify the void created and entrapped during LCM processes. The experimental results obtained in the T-junction device show the influence of wetting behaviour on bubble size. For instance, the chemical characterization of the polyester, vinylester and epoxy resins, that are partially wetting liquids, is a significant way to enhance LCM processes by improving the knowledge of the adhesion fibre/resin. Furthermore, we have proposed an analytic approach of void prediction based on both the Lucas-Washburn equation and the supplying principle for the imbibition case through an original PDM combining continuous and node interconnectivities.

**Acknowledgements**
The authors gratefully acknowledge support of this work by the Agence Nationale de Recherche under the PRC Composite project/Safran group, the LCM3M project, the CNRS and the Région Haute-Normandie.